\documentclass[conference]{IEEEtran}
\IEEEoverridecommandlockouts
\usepackage{cite}
\usepackage{amsmath,amssymb,amsfonts}
\usepackage{algorithmic}
\usepackage{graphicx}
\usepackage{textcomp}
\usepackage{xcolor}
\def\BibTeX{{\rm B\kern-.05em{\sc i\kern-.025em b}\kern-.08em
    T\kern-.1667em\lower.7ex\hbox{E}\kern-.125emX}}

\makeatletter
\newcommand{\linebreakand}{%
  \end{@IEEEauthorhalign}
  \hfill\mbox{}\par
  \mbox{}\hfill\begin{@IEEEauthorhalign}
}
\makeatother

\usepackage{amsmath}
\usepackage{comment}
\usepackage{graphicx}
\usepackage{multirow}
\usepackage{booktabs}
\usepackage{caption}
\usepackage{capt-of}  
\usepackage{array}
\usepackage{hyperref}

\usepackage{tcolorbox}
\usepackage{fontawesome5}

\usepackage{enumitem}
\setlist[itemize]{leftmargin=*} 
\usepackage{CJKutf8}

\newtcolorbox{promptbox}[2][]{
  colback=white!98!blue!2, 
  colframe=blue!70!black, 
  coltitle=white,          
  fonttitle=\bfseries\sffamily,
  title={\faTasks[regular]\hspace{1mm}~#2},
  left=2mm, right=2mm, top=1mm, bottom=1mm,
  arc=3mm,                  
  boxrule=0.9pt,
}

\usepackage{tabularx}
\usepackage{xcolor}
\usepackage{colortbl}

\newcolumntype{H}{>{\setbox0=\hbox\bgroup}c<{\egroup}@{}}

\definecolor{thedarkblue}{RGB}{0,0,120} 
\definecolor{mygreen}{RGB}{0,100,0} 
\definecolor{mydarkblue}{rgb}{0,0.08,0.45} 
\definecolor{darkblue}{rgb}{0,0.08,180}
\definecolor{orange}{HTML}{E6550D}
\colorlet{TufteRed}{red!80!black}

\definecolor{theblue}{RGB}{0,0,180}
\colorlet{thered}{TufteRed}

\providecommand{\eat}[1]{\ignorespaces}
\providecommand{\addressed}[1]{\ignorespaces}
\providecommand{\todoeat}[1]{\ignorespaces}
\providecommand{\supp}[1]{\ignorespaces} 

\DeclareMathOperator{\hugeE}{\mbox{\huge\raise-0.3ex\hbox{E}}}
\DeclareMathOperator{\p}{\mathbb{P}}
\DeclareMathOperator{\hugep}{\mbox{\huge\raise-0.3ex\hbox{$\p$}}}

\begin{document}

\title{Segment Length Matters: A Study of Segment Lengths on Audio Fingerprinting Performance}

\author{
\IEEEauthorblockN{Ziling Gong}
\IEEEauthorblockA{\textit{Data Science Institute} \\
\textit{Columbia University}\\
New York, USA \\
zg2532@columbia.edu}
\and
\IEEEauthorblockN{Yunyan Ouyang}
\IEEEauthorblockA{\textit{Data Science Institute} \\
\textit{Columbia University}\\
New York, USA \\
yo2384@columbia.edu}
\and
\IEEEauthorblockN{Iram Kamdar}
\IEEEauthorblockA{\textit{Data Science Institute} \\
\textit{Columbia University}\\
New York, USA \\
ik2594@columbia.edu}
\and
\IEEEauthorblockN{Melody Ma}
\IEEEauthorblockA{\textit{Data Science Institute} \\
\textit{Columbia University}\\
New York, USA \\
ym3065@columbia.edu}

\and
\IEEEauthorblockN{Hongjie Chen}
\IEEEauthorblockA{
\textit{Dolby Laboratories}\\
Atlanta, USA \\
0000-0002-8755-2099}

\linebreakand 

\IEEEauthorblockN{Franck Dernoncourt*}
\IEEEauthorblockA{
\textit{Adobe Research}\\
Seattle, USA \\
dernonco@adobe.com}

\and
\IEEEauthorblockN{Ryan A. Rossi*}
\IEEEauthorblockA{
\textit{Adobe Research}\\
San Jose, USA \\
ryrossi@adobe.com}

\and
\IEEEauthorblockN{Nesreen K. Ahmed}
\IEEEauthorblockA{
\textit{Cisco Research}\\
San Jose, USA \\
nesahmed@cisco.com}

\thanks{*Acted in an advisor capacity only; did not process, store, or direct the use of project data.}
}

\maketitle

\begin{abstract}
Audio fingerprinting provides an identifiable representation of acoustic signals, which can be later used for identification and retrieval systems. 
To obtain a discriminative representation, the input audio is usually segmented into shorter time intervals, allowing local acoustic features to be extracted and analyzed.
Modern neural approaches typically operate on short, fixed-duration audio segments, yet the choice of segment duration is often made heuristically and rarely examined in depth. 
In this paper, we study how segment length affects audio fingerprinting performance. 
We extend an existing neural fingerprinting architecture to adopt various segment lengths and evaluate retrieval accuracy across different segment lengths and query durations. 
Our results show that short segment lengths (0.5-second) generally achieve better performance.
Moreover, we evaluate LLM capacity in recommending the best segment length, which shows that GPT-5-mini consistently gives the best suggestions across five considerations among three studied LLMs.
Our findings provide practical guidance for selecting segment duration in large-scale neural audio retrieval systems.
\end{abstract}


\section{Introduction}
Audio segmentation is widely adopted in various audio tasks, including audio fingerprinting, speaker diarization, keyword spotting, speech translation, among others~\cite{theodorou2014overview,tsiamas22_interspeech,plaquet2025mamba,teo2025self,lee2024lightweight}.
While each task has a conventional range of segment lengths that subsequent work typically follows, the widespread use of different settings implicitly reflects that segment length is a determining factor in the task performance. 
For example, speaker identification commonly uses 3-second segments~\cite{nagrani2017voxceleb}. 
Keyword spotting models operate on segments of 0.5–1.0 seconds~\cite{rybakov2020streaming,zhang2017hello}.
Speaker diarization systems typically adopt 1.5–3 seconds to avoid multi-speaker overlap while at the same time preserving speaker-discriminative information~\cite{zhang2024variable,plaquet2025mamba,hogg2021multichannel}.
In the task of ASR, the segmentation exists as context windows, which are often 5-30 seconds but can be up to hours long~\cite{wang2025speaker,flynn2023much}.
Moreover, segmentation for generating speech language tokens has also been explored~\cite{kando2025exploring}.
On the task of audio fingerprinting, however, it remains unexplored how segment length affects model performance.

\begin{figure}[t]
\centering
\includegraphics[width=\linewidth]{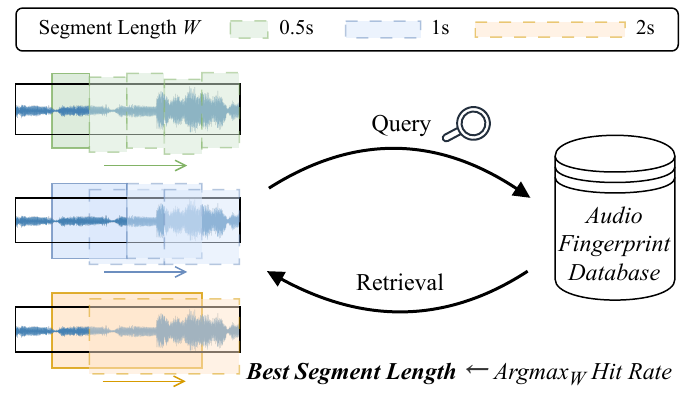}
\caption{
Our motivation question: \textit{What is the optimal segment length for the audio fingerprinting task?}
}
\label{fig:contribution-figure-demo}
\end{figure}

To bridge the gap, this paper investigates \textit{how different segment lengths impact audio fingerprinting performance}, as depicted in Fig.~\ref{fig:contribution-figure-demo}.
Specifically, we consider audio fingerprinting in the context of audio identification, where a large collection of audio recordings are segmented and fingerprinted to construct a database.
In the following query stage, a query audio is segmented and fingerprinted.
The resulting fingerprints are used to retrieve the closest segments from the database, with the goal of identifying the original version of the query audio.
We experiment with a popular neural audio fingerprinter, \textit{NAFP}~\cite{chang2021neural},
which applies 1-second segmentation to audio in both training and query stages.
In order to adopt different segment lengths, we propose a NAFP variant, named NAFP\textsuperscript{+}.
We design four sets of experiments, including a NAFP reproduction and three other experiments using $\{0.5, 1, 2\}$-second segment lengths with NAFP\textsuperscript{+}.
For each experiment set, we train an individual model with a different segment length and test with various query lengths.
Our results show that for short query lengths (shorter than 3 seconds), small segment lengths (0.5 seconds) achieve significantly better performance.
Moreover, we investigate the capacity of Large Language Models (LLMs) in suggesting segment lengths for audio fingerprinting.
We evaluate three LLMs, namely, \textit{GPT-5-mini}, \textit{Gemini-2.5-flash}, and \textit{Claude-Sonnet-4.5}.
We carefully design five prompts and collect responses from LLMs, which show that GPT-5-mini suggests the best segment length for audio fingerprinting.

To summarize, this paper has two key findings: (1) short segment length yields better audio fingerprinting performance and (2) GPT-5-mini provides the best segment length suggestions among the three studied LLMs.

\medskip\noindent\textit{\textbf{Related Work.}}
Audio fingerprinting aims to generate a compact and distinctive representation of an audio signal that enables efficient identification or retrieval from large databases.
Traditional fingerprinting systems rely on handcrafted features, such as spectral peaks~\cite{mccallum2018foreground}, while recent deep learning approaches directly extract embeddings from audio signals~\cite{su2024amgembedding,bhattacharjee2025grafprint,singh2022attention}.
For example, a recent Neural Audio FingerPrinter (NAFP) model leverages contrastive learning to generate segment embeddings~\cite{chang2021neural,fujita2024audio}.
A following work builds on top of it with a segment aggregation strategy, resulting in coarse-level segments~\cite{su2024amgembedding}.
The coarser chunk corresponds to tens of seconds (15–30 seconds), yielding fewer embeddings per track.
This saves storage and retrieval time, however, at the cost of lower accuracy.

Another work focuses on improving neural fingerprinting robustness through systematic training practices using large degraded datasets~\cite{araz2025enhancing}. 
Their results underscore the influence of training methodology on retrieval accuracy given fixed segmentation.
However, these papers have not investigated the impact of different segment lengths, which motivates our study on how different segment lengths may impact the performance of audio fingerprinting.

\begin{table}[t]
\centering
\caption{Summary of notation}

\resizebox{\linewidth}{!}{
\begin{tabular}{cl|cl}
\toprule
\textbf{Symbol} & \textbf{Description} & \textbf{Symbol} & \textbf{Description} \\
\midrule 
$a$ & audio & $f$ & sampling rate \\
$x$ & mel spectrogram & $T$ & \#STFT frames \\
$d$ & fingerprint dimension & $W$ & segment length \\
$w_{stft}$ & STFT window size & $L$ & query length \\
$h$ & hop size & $S$ & \#query segments \\
$D$ & whole audio dataset & $Q$ & query set \\
$D_{ref}$ & reference set & $D_{dist}$ & distractor set \\
\bottomrule
\end{tabular}
}
\label{tab:notation}
\end{table}

\begin{table*}[t]
\centering
\caption{
Hit Rate (\%) across Segment Lengths $W$ and Query Lengths $L$.
Bold indicates the best result within each metric group for a fixed query length $L$.
Win reports the number of query lengths where a method achieves the best performance.
}
\footnotesize
\resizebox{\linewidth}{!}{
\begin{tabular}{c c c *{11}{c} r}
\toprule
\multirow{2}{*}{\textbf{Metric}} &
\multirow{2}{*}{\textbf{Model}} &
\multirow{2}{*}{\shortstack{\textbf{Segment}\\\textbf{Length $W$}}} &
\multicolumn{11}{c}{\textbf{Query Length $L$}} &
\multirow{2}{*}{\textbf{Win}} \\
\cmidrule(lr){4-14}
& & &
\textbf{0.5} & \textbf{1} & \textbf{2} & \textbf{3} & \textbf{4} &
\textbf{5} & \textbf{6} & \textbf{7} & \textbf{8} & \textbf{9} & \textbf{10} & (\#/10) \\
\midrule

\multirow{4}{*}{\shortstack{Top1\\Exact}}
& NAFP & 1   &   -  & 73.05 & 90.60 & 95.76 & 97.35 & 98.45 & 99.00 & 99.30 & 99.45 & 99.65 & 99.70 & - \\
& \multirow{3}{*}{\rotatebox{45}{NAFP\textsuperscript{+}}}
& 0.5 & 68.50 & \textbf{85.70} & \textbf{94.30} & \textbf{96.70} & \textbf{97.95} & \textbf{98.75} & \textit{98.90} & \textbf{99.35} & \textbf{99.55} & \textbf{99.65} & 99.65 & \textbf{8.5 / 10} \\
& & 1   & - & 75.55 & 91.65 & 96.40 & 97.70 & 98.45 & \textit{98.90} & 99.20 & 99.40 & 99.60 & \textbf{99.70} & 1.5 / 10 \\
& & 2   & - & - & 72.05 & 85.45 & 89.60 & 92.00 & 93.20 & 94.15 & 95.15 & 95.40 & 95.60 & 0 / 10 \\
\midrule

\multirow{4}{*}{\shortstack{Top3\\Exact}}
& NAFP   & 1   &   -  & 81.00 & 93.47 & 96.90 & 98.45 & 99.00 & 99.30 & 99.50 & 99.60 & 99.75 & 99.75 & - \\
& \multirow{3}{*}{\rotatebox{45}{NAFP\textsuperscript{+}}}
& 0.5 & 75.45 & \textbf{89.10} & \textbf{95.80} & \textbf{97.80} & \textbf{98.60} & \textbf{99.05} & \textbf{99.30} & \textbf{99.55} & \textbf{99.70} & \textbf{99.75} & \textit{99.75} & \textbf{9.5 / 10} \\
& & 1   &   -  & 81.65 & 93.45 & 97.40 & 98.45 & 98.80 & 99.15 & 99.35 & 99.55 & 99.65 & \textit{99.75} & 0.5 / 10 \\
& & 2   &   -  &   -   & 78.25 & 87.75 & 91.65 & 93.55 & 94.50 & 95.05 & 95.75 & 95.95 & 96.00 & 0 / 10\\
\midrule

\multirow{4}{*}{\shortstack{Top10\\Exact}}
& NAFP   & 1   &   -  & 84.10 & 96.50 & 97.55 & 98.80 & 99.25 & 99.35 & 99.50 & 99.60 & 99.80 & 99.80 & - \\
& \multirow{3}{*}{\rotatebox{45}{NAFP\textsuperscript{+}}}
& 0.5 & 78.80 & \textbf{90.40} & \textbf{96.45} & \textbf{98.00} & \textbf{98.75} & \textbf{99.20} & \textbf{99.55} & \textbf{99.80} & \textbf{99.85} & \textbf{99.90} & \textbf{99.85} & \textbf{10 / 10} \\
& & 1   & - & 84.40 & 94.65 & 97.65 & 98.70 & 99.00 & 99.30 & 99.45 & 99.55 & 99.70 & 99.75 & 0 / 10 \\
& & 2   & - & - & 81.05 & 89.15 & 92.55 & 93.90 & 94.80 & 95.30 & 95.85 & 95.95 & 96.05 & 0 / 10 \\
\midrule
\midrule
\multirow{4}{*}{\shortstack{Top1\\Near}}
& NAFP   & 1   &   -  & 78.00 & 91.70 & 96.30 & 97.55 & 98.55 & 99.00 & 99.30 & 99.50 & 99.65 & 99.70 & -\\
& \multirow{3}{*}{\rotatebox{45}{NAFP\textsuperscript{+}}}
& 0.5 & \textbf{72.10} & \textbf{87.20} & \textbf{94.70} & \textbf{96.80} & \textbf{98.00} & \textbf{98.75} & \textbf{98.95} & \textbf{99.35} & \textbf{99.55} & \textbf{99.65} & 99.65 & \textbf{9 / 10} \\
& & 1   & - & 78.35 & 92.35 & 96.40 & 97.70 & 98.45 & 98.90 & 99.25 & 99.45 & 99.60 & \textbf{99.70} & 1 / 10 \\
& & 2   & - & - & 76.25 & 86.35 & 90.15 & 92.35 & 93.55 & 94.35 & 95.20 & 95.50 & 95.70 & 0 / 10 \\

\bottomrule
\end{tabular}
}
\label{tab:hit-rate-comparison}
\end{table*}

\begin{figure*}[t]
    \centering
    \includegraphics[width=1\linewidth]{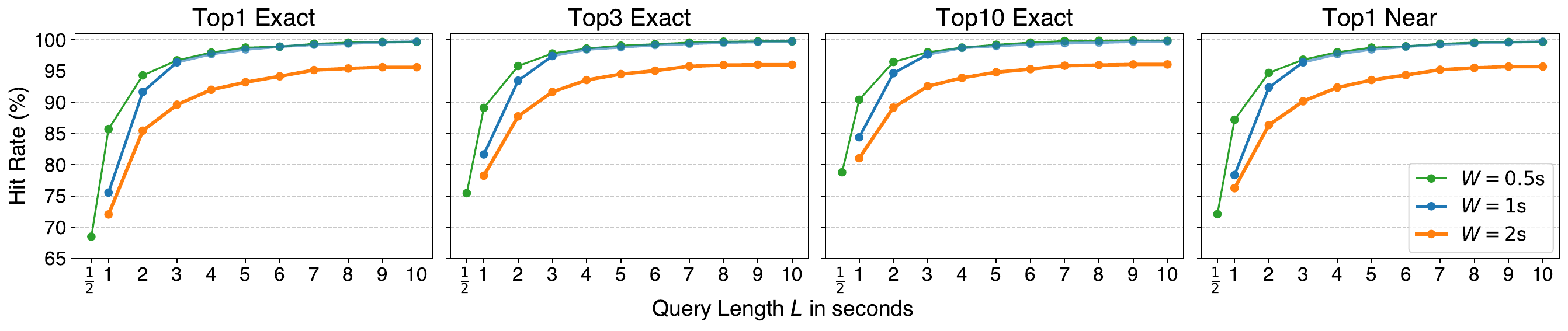}
    \caption{
        Hit Rate (\%) of Various Segment Lengths $W$ (Green: $W=0.5$, Blue: $W=1$, Orange: $W=2$) and Query Lengths $L$.
    } 
    \label{fig:hr-vs-querylen}
\end{figure*}

\section{Experiments with NAFP and NAFP\textsuperscript{+}}
This section describes the experimental setups with both the original NAFP and our proposed model variant NAFP\textsuperscript{+}.

\subsection{Preliminary}
Let $a$ denote an audio recording and $f$ its sampling rate.
We use $W$ to denote the segment length.
For evaluation, we let $D = D_{ref} \cup D_{dist}$ denote a collection of audio recordings, where $D_{ref}$ denotes a reference set that derives an audio query set $Q$, while $D_{dist}$ denotes a distractor set providing unrelated candidates.
Each audio $q \in Q$ is queried with a query length $L$.
A moving window size of $W$ and hop of $h$ is applied to this $L$-second (sub)audio to generate query segments.
Let $S$ denote the number of resulting query segments, hence, $q_1, q_2, \ldots, q_S$. 

\subsection{NAFP\textsuperscript{+} Design}
Given an audio $a$, we apply a sliding window size of $W \in \{0.5, 1, 2\}$ seconds and a hop of $h=0.5$ seconds to obtain training segments.
We then apply mel-spectrogram transformation to each segment, resulting in an input $x\in \mathbb{R}^{F \times T}$,
where $F = 256$ is a user-selected number of Mel-frequency bins.
$T$ is the number of resulting frames from Short-Time Fourier Transform with window $w_{stft}=256$,
\begin{align}
T = \lfloor \frac{W \cdot f}{w_{stft}} \rfloor + 1 = \lfloor \frac{W \cdot 8000}{256} \rfloor + 1
\end{align}
Specifically, when $W$ is $0.5$, $1$, and $2$ seconds, the corresponding $T$ is $16$, $32$, and $63$.
For NAFP, the mel spectrogram $x\in \mathbb{R}^{F \times T}$ goes through a chain of eight convolution blocks. 
Each block consists of a temporal convolutional layer, a frequency convolution layer, ELU activation, and a normalization layer.
L2-normalization is applied to the final fingerprint of a selected dimension ($d=128$).
We follow the same augmentation and training scheme as NAFP~\cite{chang2021neural}, where an audio and its distorted replica are considered a positive pair in contrastive learning.

In order to adopt segment lengths other than 1-second, we propose a model variant named NAFP\textsuperscript{+}.
To minimize changes to the convolutional blocks in NAFP, we derive NAFP\textsuperscript{+} by adding a fully-connected layer with ELU activation before each block.
This modification normalizes the number of audio frames from $T$ to a fixed number $T_0=32$.

\begin{table*}[t]
\centering
\small
\footnotesize
\renewcommand{\arraystretch}{1.0}
\caption{
LLM-recommended segment lengths across 5 prompts across different LLMs.
}
\resizebox{\linewidth}{!}{
\begin{tabular}{p{1pt}p{4cm}p{4cm}p{4.1cm}p{4cm}}
\toprule
\textbf{\#} & \textbf{Prompt} & \textbf{GPT-5-mini}& \textbf{Gemini-2.5-flash} & \textbf{Claude-Sonnet-4.5}\\
\midrule
1 & What segment duration best balances temporal resolution and discriminative power in this setup? & Pick a segment duration in the \textbf{1–2 second} range. It gives a good tradeoff between temporal resolution and discriminative power for music fingerprinting & Optimal Range: \textbf{3-5 seconds}. This duration is typically long enough to capture meaningful musical phrases, harmonic progressions, and rhythmic grooves that define a unique musical identity. & The optimal segment duration for audio fingerprinting typically falls between 1-3 seconds, with many systems using around \textbf{2 seconds} as a good balance point. \\
\cmidrule(lr){2-5}
2 & When audio segments are affected by background noise and small time shifts, what segment duration yields the most stable fingerprints without losing detail? & Aim for about \textbf{1 second} per segment (a practical range ~0.8–1.5 s, with 1.0 s the usual sweet spot). & A segment duration of \textbf{2 to 3 seconds} is often a good starting point, providing a strong balance between stability, detail, and computational efficiency. & A segment duration of around \textbf{1-2 seconds} typically provides the best balance between fingerprint stability and temporal resolution.\\
\cmidrule(lr){2-5}
3 & In a large-scale contrastive audio fingerprinting system, what segment duration provides the best trade-off between retrieval accuracy, speed, and storage cost? & \textbf{1 second} is the best default trade-off for a large-scale, high-specificity contrastive audio‑fingerprinting system. & \textbf{3 to 5 seconds} is a strong starting point. This range offers a good compromise between capturing sufficient musical context, managing storage costs, and maintaining acceptable indexing speeds. & The optimal segment duration in a contrastive audio fingerprinting system typically falls between 1-3 seconds, with many systems using around \textbf{2 seconds} as a good compromise.\\
\cmidrule(lr){2-5}
4 & In a large-scale contrastive system, what duration provides the best trade-off between accuracy, speed, and storage cost? & Use segments in the \textbf{~1–2 second range} (practical sweet spot about 1.0–1.5 s; 2 s if you prefer fewer embeddings/storage at a small accuracy cost). & A segment duration in the range of \textbf{2 to 5 seconds} with 50\% overlap often provides the best trade-off. & The optimal segment duration in contrastive neural audio fingerprinting typically falls between \textbf{1-3 seconds}, with many systems using 1.5-2 seconds.\\
\cmidrule(lr){2-5}
5 & When training with fixed-duration segments but testing with variable-duration queries, which segment duration generalizes best across query lengths? & 
Short, roughly \textbf{1‑second} segments (or at most ~1–2 s) give the best overall generalization when you train with a single fixed segment length but must handle variable‑length queries. & \textbf{3 to 8 seconds} is a commonly cited and empirically supported sweet spot, offering the best generalization across a range of query lengths due to its balance of musical context and flexibility. & The \textbf{1-2s} range typically provides the best generalization across varying query lengths while maintaining practical efficiency.\\
\midrule
\multicolumn{2}{l}{Summary} & \textbf{\textit{1-2 seconds}} (consistent) & \textbf{\textit{2–8 seconds}} (fluctuating) &  \textbf{\textit{1–3 seconds}} (centered at 2 secs)\\
\bottomrule
\end{tabular}
}
\vspace{-1mm}
\label{tab:llm_segment_length_b}
\end{table*}

\subsection{Evaluation}
We evaluate model performance on an audio identification task.
We consider an audio dataset $D=D_{ref}\cup D_{dist}$ and partition it into a reference set $D_{ref}$ and a distractor set $D_{dist}$.
The reference set $D_{ref}$ is used to derive a query set $Q$, while $D_{dist}$ serves as unrelated candidates to prevent the retrieval task from being trivial.
Our objective is to identify the correct original audio of $q \in Q$ from $D$, particularly from $D_{ref}$.

For a query audio, instead of using all of its audio frames, a \textit{query length} of $L$ seconds is applied to each query audio $q$ to select a (sub)audio for the query.
The selected portion is further segmented with $W$-second windows and a $h$-second hop size, resulting in $S$ segments where
\begin{align}
S = \left\lfloor \frac{L - W}{h} \right\rfloor + 1.
\label{eq:segment_length}
\end{align}
For instance, if the query length is $L = 3$ seconds, there are $S = {(3 - 1)} / {0.5}+1 = 5$ segments, with $W = 1$ and $h = 0.5$ seconds.
For each of the resulting $S$ segments, namely, $q_1, q_2, \ldots, q_S$, we retrieve its closest segments and retrieve the corresponding audio from weighted segment majority voting.

\begin{figure}[!ht]
\centering
\footnotesize
\begin{promptbox}{Prompt Template}
\begin{small}
\vspace{-1mm}

\noindent \parbox { \linewidth } {
\texttt{Segment duration refers to the duration of each fixed-length audio segment measured in seconds in a contrastive neural audio fingerprinting model designed for high-specific retrieval. The model generates one embedding per segment, which is later used for similarity-based matching in a large-scale music database.}

\texttt{\texttt{ }}

\texttt{Dataset context: Training uses 30-second clips from the Free Music Archive (FMA) dataset with strict train, validation, and test splits. Evaluation involves large-scale retrieval, where each track in the database is segmented into fixed windows and the query is matched by nearest-neighbor search in the embedding space.}

\texttt{\texttt{ }}

\texttt{Example: Consider 1-second segments with a 0.5-second hop capturing a short melodic phrase or drum pattern used to identify its source track in a large database under background noise and small time shifts.}

\texttt{\texttt{ }}

\texttt{\{Question Prompt\}}

}
\end{small}
\vspace{-2mm}
\end{promptbox}
\caption{Prompt Template: Fixed Context Prompt + One of the Five Question Prompts from Table~\ref{tab:llm_segment_length_b}.}
\label{fig:prompt-template}
\end{figure}

\medskip\noindent\textbf{\textit{Dataset.}}
We use a derived music dataset from the Free Music Archive collection~\cite{defferrard2016fma,chang2021neural}.
Each audio in the dataset is a musical piece that lasts for 30 seconds.
The dataset is partitioned into three sets:
A training set with 10,000 audio clips (83.3 hours), 
a reference audio set $D_{ref}$ with 500 audio clips (4.2 hours),
and a distractor set $D_{dist}$ with 9,978 audio clips (83.3 hours).
A query set $Q$ is derived from the reference set by 
synthesizing with a held-out subset of background noises, augmenting with time offsets, and simulating of room impulse responses.

\medskip\noindent\textbf{\textit{Metrics.}}
We measure \textit{Top-K Exact Hit}, which considers whether the correct retrieval is within top-K predictions, and \textit{Top-K Near Hit}, which allows an extra $\pm1$ frame misalignment.
We report the Top-K Hit Rate across query segments, i.e., average Top-K-Exact Hit for $K\in \{1, 3, 10\}$ and Top-1-Near Hit.
Higher Hit Rate means better performance and hence more favorable.

\section{Results}
We report results of NAFP and NAFP\textsuperscript{+} on segment lengths $W \in \{0.5, 1, 2\}$ seconds and query lengths $L \in \{1, 2, \ldots, 10\}$ seconds, as shown in Table~\ref{tab:hit-rate-comparison}.
We have multiple observations:
(1) Segment length $W=0.5$ seconds achieves the best performance on most query lengths, with a higher than 8 / 10 win rate in any metric group.
In contrast, $W=2$ seconds has the worst performance across query lengths.
(2) For small query lengths ($L \le 3$), shorter segment lengths consistently achieve significantly higher Hit Rates across metrics.
As query length $L$ increases beyond 3 seconds, the performance gap between $W=0.5$ seconds and $W=1$ second narrows substantially.
This suggests that additional temporal context compensates for coarser segmentation.
(3) We further visualize the results in Fig.~\ref{fig:hr-vs-querylen}.
The figure shows that performance improves rapidly with additional segments and begins to saturate after 4 seconds, and only very slight improvement afterwards.

\subsection{Segment Length Selection with LLMs}
We examine the capacity of LLMs in suggesting the best segment lengths, i.e., the segment length that yields the highest hit rates.
We evaluate three models, namely, \textit{GPT-5-mini}, \textit{Gemini-2.5-flash}, and \textit{Claude-Sonnet-4.5}, with five different prompts.
Each prompt is constructed as a combination of a fixed context prompt and a question prompt.
The fixed context prompt includes a baseline task description, dataset context based on FMA and large-scale retrieval, and a concrete segmentation example illustrating short overlapping segments under background noise and small time shifts, as depicted in Fig.~\ref{fig:prompt-template}.
The question prompts are designed from five considerations, including
(1) balancing temporal resolution and discriminative power,
(2) stability under background noise and small time shifts,
(3) the trade-off between retrieval accuracy, speed, and storage cost in large-scale systems,
(4) a simplified formulation of the accuracy--speed--storage trade-off, and
(5) generalization when training on fixed-duration segments but testing on variable-duration queries.
The corresponding question wording prompts are provided in Table~\ref{tab:llm_segment_length_b}.

We compare LLM recommendations across the five question prompts and observe that Gemini exhibits higher run-to-run variability under the same prompt.
In contrast, GPT-5-mini consistently recommends a segment length of approximately 1 second across different question wordings,
whereas Gemini and Claude more frequently suggest longer durations around or above 2 seconds.
When compared with our empirical findings, the 1-second recommendation aligns more closely with the optimal accuracy, which indicates that GPT-5-mini gives the best segment length recommendations.

\section{Conclusion}
In this paper, we investigate the impact of different segment lengths on the performance of the audio fingerprinting task.
The performance gain saturates after 4-second queries regardless of the selected segment lengths. 
We further investigate LLM capacity in suggesting segment lengths with well-designed prompts.
GPT-5-mini consistently recommends a 1-second segment length across different question prompts, aligning closely with our empirical results, whereas Gemini-2.5-flash and Claude-Sonnet-4.5 exhibit higher variability and tend to suggest longer durations.
These results indicate that appropriately contextualized LLMs reflect empirically grounded design choices, though their reliability varies across models.

Future research directions include examining whether these findings generalize to other audio domains beyond music, such as speech or environmental sounds.
Moreover, segment lengths may also impact other audio tasks, such as speaker verification, keyword spotting, etc., which can be further explored.
Finally, research potential also lies in developing automated strategies for segment length selection based on query length or audio content, with LLMs or lightweight predictors as  tools.
\bibliographystyle{IEEEtran}
\bibliography{main}

@article{theodorou2014overview,
  title={An overview of automatic audio segmentation},
  author={Theodorou, Theodoros and Mporas, Iosif and Fakotakis, Nikos},
  journal={International Journal of Information Technology and Computer Science (IJITCS)},
  volume={6},
  number={11},
  pages={1},
  year={2014}
}

@inproceedings{chang2021neural,
  title={Neural audio fingerprint for high-specific audio retrieval based on contrastive learning},
  author={Chang, Sungkyun and Lee, Donmoon and Park, Jeongsoo and Lim, Hyungui and Lee, Kyogu and Ko, Karam and Han, Yoonchang},
  booktitle={ICASSP 2021-2021 IEEE International Conference on Acoustics, Speech and Signal Processing (ICASSP)},
  pages={3025--3029},
  year={2021},
  organization={IEEE}
}

@inproceedings{bhattacharjee2025grafprint,
  title={Grafprint: A gnn-based approach for audio identification},
  author={Bhattacharjee, Aditya and Singh, Shubhr and Benetos, Emmanouil},
  booktitle={ICASSP 2025-2025 IEEE International Conference on Acoustics, Speech and Signal Processing (ICASSP)},
  pages={1--5},
  year={2025},
  organization={IEEE}
}

@article{singh2022attention,
  title={Attention-based audio embeddings for query-by-example},
  author={Singh, Anup and Demuynck, Kris and Arora, Vipul},
  journal={arXiv preprint arXiv:2210.08624},
  year={2022}
}

@article{araz2025enhancing,
  title={Enhancing Neural Audio Fingerprint Robustness to Audio Degradation for Music Identification},
  author={Araz, R Oguz and Cortes-Sebastia, Guillem and Molina, Emilio and Serra, Joan and Serra, Xavier and Mitsufuji, Yuki and Bogdanov, Dmitry},
  journal={arXiv preprint arXiv:2506.22661},
  year={2025}
}

@inproceedings{
su2024amgembedding,
title={{AMG}-Embedding: a Self-Supervised Embedding Approach for Audio Identification},
author={Yuhang Su and Wei Hu and Fan Zhang and Qiming Xu},
booktitle={ACM Multimedia 2024},
year={2024},
url={https://openreview.net/forum?id=H7etFJugLW}
}

@article{defferrard2016fma,
  title={FMA: A dataset for music analysis},
  author={Defferrard, Micha{\"e}l and Benzi, Kirell and Vandergheynst, Pierre and Bresson, Xavier},
  journal={arXiv preprint arXiv:1612.01840},
  year={2016}
}

@article{nagrani2017voxceleb,
  title={Voxceleb: a large-scale speaker identification dataset},
  author={Nagrani, Arsha and Chung, Joon Son and Zisserman, Andrew},
  journal={arXiv preprint arXiv:1706.08612},
  year={2017}
}

@article{zhang2017hello,
  title={Hello edge: Keyword spotting on microcontrollers},
  author={Zhang, Yundong and Suda, Naveen and Lai, Liangzhen and Chandra, Vikas},
  journal={arXiv preprint arXiv:1711.07128},
  year={2017}
}

@article{rybakov2020streaming,
  title={Streaming keyword spotting on mobile devices},
  author={Rybakov, Oleg and Kononenko, Natasha and Subrahmanya, Niranjan and Visontai, Mirk{\'o} and Laurenzo, Stella},
  journal={arXiv preprint arXiv:2005.06720},
  year={2020}
}

@inproceedings{zhang2024variable,
  title={Variable Segment Length and Domain-Adapted Feature Optimization for Speaker Diarization},
  author={Zhang, Chenyuan and Luo, Linkai and Peng, Hong and Wen, Wei},
  booktitle={Proc. Interspeech 2024},
  pages={3744--3748},
  year={2024}
}

@article{flynn2023much,
  title={How much context does my attention-based ASR system need?},
  author={Flynn, Robert and Ragni, Anton},
  journal={arXiv preprint arXiv:2310.15672},
  year={2023}
}

@article{wang2025speaker,
  title={Speaker Targeting via Self-Speaker Adaptation for Multi-talker ASR},
  author={Wang, Weiqing and Park, Taejin and Medennikov, Ivan and Wang, Jinhan and Dhawan, Kunal and Huang, He and Koluguri, Nithin Rao and Balam, Jagadeesh and Ginsburg, Boris},
  journal={arXiv preprint arXiv:2506.22646},
  year={2025}
}

@inproceedings{plaquet2025mamba,
  title={Mamba-based segmentation model for speaker diarization},
  author={Plaquet, Alexis and Tawara, Naohiro and Delcroix, Marc and Horiguchi, Shota and Ando, Atsushi and Araki, Shoko},
  booktitle={ICASSP 2025-2025 IEEE International Conference on Acoustics, Speech and Signal Processing (ICASSP)},
  pages={1--5},
  year={2025},
  organization={IEEE}
}

@inproceedings{tsiamas22_interspeech,
  author={Ioannis Tsiamas and Gerard I. Gállego and José A. R. Fonollosa and Marta R. Costa-jussà},
  title={{SHAS: Approaching optimal Segmentation for End-to-End Speech Translation}},
  year=2022,
  booktitle={Proc. Interspeech 2022},
  pages={106--110},
  doi={10.21437/Interspeech.2022-59}
}

@inproceedings{teo2025self,
  title={Self-Information Guided Speech Segmentation for Efficient Streaming ASR},
  author={Teo, Wen Shen and Minami, Yasuhiro},
  booktitle={ICASSP 2025-2025 IEEE International Conference on Acoustics, Speech and Signal Processing (ICASSP)},
  pages={1--5},
  year={2025},
  organization={IEEE}
}

@inproceedings{hogg2021multichannel,
  title={Multichannel overlapping speaker segmentation using multiple hypothesis tracking of acoustic and spatial features},
  author={Hogg, Aidan OT and Evers, Christine and Naylor, Patrick A},
  booktitle={ICASSP 2021-2021 IEEE International Conference on Acoustics, Speech and Signal Processing (ICASSP)},
  pages={26--30},
  year={2021},
  organization={IEEE}
}

@article{lee2024lightweight,
  title={Lightweight Audio Segmentation for Long-form Speech Translation},
  author={Lee, Jaesong and Kim, Soyoon and Kim, Hanbyul and Chung, Joon Son},
  journal={arXiv preprint arXiv:2406.10549},
  year={2024}
}

@article{kando2025exploring,
  title={Exploring the Effect of Segmentation and Vocabulary Size on Speech Tokenization for Speech Language Models},
  author={Kando, Shunsuke and Miyao, Yusuke and Takamichi, Shinnosuke},
  journal={arXiv preprint arXiv:2505.17446},
  year={2025}
}

@article{fujita2024audio,
  title={Audio Fingerprinting with Holographic Reduced Representations},
  author={Fujita, Yusuke and Komatsu, Tatsuya},
  journal={arXiv preprint arXiv:2406.13139},
  year={2024}
}

@inproceedings{mccallum2018foreground,
  title={Foreground Harmonic Noise Reduction for Robust Audio Fingerprinting},
  author={McCallum, Matthew C},
  booktitle={2018 IEEE International Conference on Acoustics, Speech and Signal Processing (ICASSP)},
  pages={3146--3150},
  year={2018},
  organization={IEEE}
}
\end{document}